\documentclass[10pt]{iopart}
\usepackage{iopams}
\usepackage{graphicx}% Include figure files
\usepackage{color}
\usepackage{subcaption}
\usepackage{float}
\usepackage{bm}

\begin{document}

\title{Thermal vestiges of avalanches in the driven random field Ising model}

\author{Liheng Yao$^1$ and Robert L. Jack$^{1,2}$}
\address{$^1$DAMTP, Centre for Mathematical Sciences, University of Cambridge,
Wilberforce Road, Cambridge CB3 0WA, United Kingdom}
%\ead{ly343@cam.ac.uk}
\address{$^2$ Yusuf Hamied Department of Chemistry, University of Cambridge, Lensfield Road, Cambridge CB2 1EW, United Kingdom}

\definecolor{mygreen}{rgb}{0.0,0.55,0.3}
\newcommand{\rlj}[1]{{\color{mygreen}#1}}

\newcommand{\ly}[1]{{\color{blue}#1}}

\newcommand{\beq}{\begin{equation}}
\newcommand{\eeq}{\end{equation}}

%\vspace{10pt}
%\begin{indented}
%\item[]July 2022
%\end{indented}

\begin{abstract}
We investigate the non-equilibrium behavior of the $3d$ random field Ising model at finite temperature, as an external field is increased through its coercive field.
We show by numerical simulations that the phenomenology of avalanches -- which are sharply defined only at zero temperature -- also persists over a significant range of finite temperatures.
We analyse the main differences between the thermal and zero-temperature systems, including an excess of small avalanches in the thermal case, whose behaviour is consistent with activated dynamical scaling.
We also investigate the extent to which individual avalanches at finite temperature can be traced back to parent avalanches in the athermal system.
\end{abstract}

%
% Uncomment for keywords
%\vspace{2pc}
%\noindent{\it Keywords}: ***
%
% Uncomment for Submitted to journal title message
%\submitto{\JPA}
%
% Uncomment if a separate title page is required
%\maketitle
% 
% For two-column output uncomment the next line and choose [10pt] rather than [12pt] in the \documentclass declaration
%\ioptwocol
%

\section{Introduction}

%intro

The random field Ising model (RFIM) is a simple spin model with quenched disorder, which has several interesting properties.  At equilibrium, it has a transition between paramagnetic and ferromagnetic states where the disorder is relevant in the renormalisation group (RG) sense \cite{grinstein_ferromagnetic_1976, bray_scaling_1985}. The resulting critical behaviour is described by a zero-temperature fixed point, and differs qualitatively from standard (disorder-free) criticality. Its phenomenology includes activated dynamics at finite temperature: the critical slowing down of the model is exponential in the correlation length, instead of the usual power law \cite{fisher_scaling_1986, villain_nonequilibrium_1984}.  The model also supports a form of non-equilibrium criticality at zero temperature, which occurs when an external magnetic field is increased \cite{sethna_hysteresis_1993, perkovic_avalanches_1995, perkovic_disorder-induced_1999}. In this case, spins flip as part of correlated avalanches, and these avalanches have scale-free behaviour near criticality~\cite{perkovic_disorder-induced_1999,sethna_hysteresis_1993}.  This behaviour can be related to crackling (Barkhausen) noise \cite{barkhausen_1919} that occurs on increasing the magnetic field in impure magnets \cite{perkovic_avalanches_1995, salje_crackling_2014, xu_barkhausen_2015}. By contrast with spin glass models {(see e.g. \cite{castellani_spin-glass_2005})}, all this behaviour takes place without any frustrated interactions: this  simplifies the theoretical descriptions.

Relaxation by correlated avalanches occurs in a variety of contexts, including amorphous systems under athermal quasistatic shear \cite{maloney_amorphous_2006} and  dense active matter \cite{morse_direct_2021, mandal_how_2021}. 
The RFIM provides a generic paradigm for such non-equilibrium dynamics.  Its behaviour under athermal driving
has been used to model a diverse range of phenomena, such as capillary condensation in disordered porous media \cite{PhysRevLett.71.4186, aubry_condensation_2014, detcheverry_mechanisms_2004}, yielding of disordered materials under stress \cite{da_silveira_introduction_1999, lin_scaling_2014}, particularly yielding of supercooled liquids under athermal quasistatic shear \cite{ozawa_random_2018}, and crises and collective phenomena in socio-economic systems \cite{bouchaud_crises_2013}. Moreover, it was shown that the dynamical transition in the mode-coupling theory of glasses is in the universality class of the critical point of the non-equilibrium RFIM, where an infinite avalanche first emerges \cite{franz_field_2011}.

The essential idea in these cases is that a locally-stable minimum on the energy landscape will eventually become unstable as some parameter (the ``driving field'') is increased.  When this happens, the system will relax its energy and find a new minimum. The associated change in the system state might be small and local, or large and long-ranged.  Such changes are identified as avalanches, because the system moves suddenly down the energy landscape.  At the non-equilibrium critical point of the RFIM, one finds avalanches of all sizes (up to the system size itself): the distribution of the avalanche sizes is scale-free.  The identification of a universal description for such phenomena is a striking discovery in statistical physics.  

From the theoretical point of view, all these universal phenomena take place at zero temperature, hinting at some generic structures in their energy landscapes.
On the other hand, physical phenomena like Barkhausen noise is observed at finite temperature, so it is natural to ask about thermal effects on the avalanche behaviour.  
Indeed, recent experimental work has found scale-free avalanche size scaling of RFIM type in an artificial spin-ice array of interacting nanomagnets \cite{bingham_experimental_2021}, and a rare-earth ferromagnet in a transverse magnetic field \cite{xu_barkhausen_2015}. Particularly for the latter case, the authors identify a breakdown of low-temperature avalanche size scaling, near room temperature.  The interplay of thermal effects with avalanches is particularly relevant for connections between RFIM physics and glassy materials, where activation effects from thermal fluctuations lead to a wealth of complex behaviour, and it is important to understand how these interact with the universal RFIM physics.

This work considers how RFIM avalanches are affected at finite temperature $T$.
In the mean-field RFIM, it is easy to see that the universal physics only survives for vanishingly small $T$.  However, the situation in  in three dimensions is less clear: avalanches are no longer sharply-defined for $T>0$ because of thermal activation, but one may still expect flipping of large correlated domains of spins in response to non-equilibrium driving.  Heuristically, one expects two competing time scales in such driven systems: the inverse driving rate $\Gamma ^{-1}$, and the equilibrium (thermal) relaxation time $\tau _{\rm eq} (T)$.  For $\Gamma \tau _{\rm eq} \gg 1$, we expect the athermal non-equilibrium behaviour to survive, while for $\Gamma \tau _{\rm eq} \ll 1$ we expect equilibrium relaxation.  We investigate the crossover between these two regimes, focussing on the extent to which avalanches can be identified in thermal systems, and what properties they have.

After defining a method to identify thermally-activated avalanches, our main qualitative result is that the zero-temperature scaling behaviour of avalanches can still be observed at relatively high temperatures $T$, of the same order as the exchange interaction energy of the model.  This observation supports the applicability of RFIM-type ideas to thermal systems like sheared glasses.  
The main corrections to the zero-temperature behaviour are (i) an excess contribution of small avalanches, which are excited reversibly by thermal activation; and (ii) a cutoff for large avalanches, which we attribute primarily to {the finite correlation length of thermal fluctuations, and the fact that the finite driving rate does not allow for the relaxation of arbitrarily large domains}. In addition, we discuss {the dependencies of the avalanche statistics on the driving protocol}, and the extent to which specific thermal avalanches can be identified with athermal ``parents'' (at fixed disorder).

%here the extent to which the universal (athermal) behaviour can be observed in thermal systems \ly{(i.e. the range for which $\Gamma \tau _{eq} \gg 1$), and the nature of the crossover towards equilibrium}.

Before embarking on this analysis, we 
note that the effects of thermal fluctuations on RFIM dynamics have been studied before, in different contexts.  The depinning transition of an infinite interface in the two-dimensional RFIM have been studied previously, and it was found that the there is a thermal rounding of the transition \cite{PhysRevE.60.5202}, and the low-temperature creep motion below the depinning threshold is asymptotically logarithmic in time \cite{PhysRevE.63.026113, Dong_2012}. The low-temperature coarsening process of the two-dimensional RFIM has also been studied \cite{PhysRevE.87.022121, PhysRevE.89.042144}, with the same asymptotic domain growth. In addition, the equilibrium criticality of the RFIM, where for sufficiently small disorder there is an $R$-dependent critical temperature $T_c$ below which the equilibrium state becomes ferromagnetic, has been investigated extensively (see e.g. \cite{doi:10.1142/9789812819437_0009} for a classic review, \cite{tarjus_random-field_2020} for a review of recent theoretical advancements, and \cite{fytas_review_2018} for a review of recent numerical work). This equilibrium criticality is distinct to the non-equilibrium criticality that concerns our work: it was shown that the two critical points are not in the same universality class in the physical dimensions that we operate in \cite{balog_criticality_2018}.

In the remainder of this paper, Sec.~\ref{sec:model} introduces the RFIM and outlines our numerical methods, and some relevant results from the literature.  Sec.~\ref{sec:results} presents our  results, and we draw our conclusions in Sec.~\ref{sec:conc}.

\section{Random field Ising model}
\label{sec:model}

\subsection{Model definition, and dynamical evolution}

The $d$-dimensional RFIM is defined on a (hyper)cubic lattice of linear size $L$.  Each site has an Ising spin $s_i$ which takes the values $\pm1$, which are the `up' state and the `down' state. There are $N=L^d$ spins in total.   Spin $i$ interacts with its neighbours by an exchange interaction of strength $J$, and it feels a magnetic field of strength $H+h_i$, so the system's energy is
\begin{equation}
    E = -J \sum_{\langle ij \rangle}s_{i} s_{j} -\sum_{i}  (H+h_i) s_{i} \; ,
    \label{ham}
\end{equation}
where the notation $\langle ij \rangle$ indicates a sum over pairs of nearest neighbours.  Here and throughout, sums over $i$ run over all spins in the system, unless explicitly stated otherwise.  The parameter $H$ represents the external magnetic field, and $h_i$ is a quenched random field on site $i$.  The fields on different sites are independent, they are Gaussian distributed with covariance $\overline{ h_{i}h_{j} } = R^{2}\delta_{ij}$, where the notation $\overline{(\cdot)}$ indicates an average over the quenched disorder.  Hence the probability density for a single field is 
%where the spins $s_{i}$ take values $\pm 1$, the index $\langle ij \rangle$ of the first summation runs over nearest neighbours, $J$ is the exchange constant which will be chosen to be $1$ for the remainder of this work, $H$ is the external field, and $h_{i}$ is a quenched random field following a zero-mean Gaussian distribution with $\langle h_{i}h_{j} \rangle = R^{2}\delta_{ij}$, so that the probability density function of the random fields is
\begin{equation}
    \rho (h_{i}) = \frac{1}{\sqrt{2 \pi R^2}} \exp \left( -\frac{h_{i}^{2}}{2 R^{2}} \right) \; .
    \label{pdf}
\end{equation}
Physically, $R$ is the typical magnitude of the random field.
We fix the energy scale by setting $J=1$, so the dimensionless parameters that appear in the energy are $H$ and $R$.

To study the RFIM at finite temperature $T$, it is natural to use Monte Carlo (MC) dynamics. In a single MC move, one picks a random spin $s_i$ and proposes to change its value from $s_i$ to $s_i$.  This proposed move is accepted with probability ${\rm min}(1,{\rm e}^{-\Delta E_{i}/T})$ where $\Delta E_{i}$ is the change in energy for the proposed move~\cite{metropolis_equation_1953}.  A sequence of $N$ such moves is called an MC sweep.  The natural (dimensionless) time unit for the dynamics is 1 MCS.

For $T=0$, running MC dynamics results in a local energy minimization, in which one flips all spins for which $\Delta E_i$ would be negative, and repeats this process until $\Delta E_i>0$ for all spins $i$.  If one or more spins is flipped during the minimization, this is called an avalanche, and the number of spins that flipped is the avalanche size $S$.

\begin{figure}[t]
    \centering
    \includegraphics[width=1\linewidth]{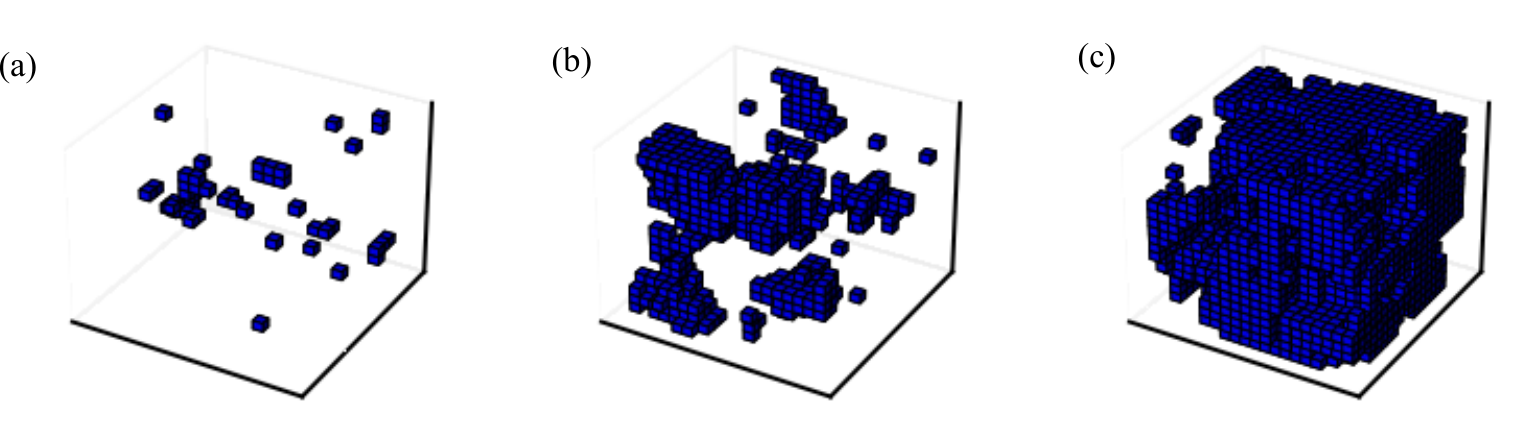}
    \caption{Snapshots of spins that have flipped in the athermal driven RFIM at the critical disorder $R = 2.2$ and (a) $H = H_{c} - 0.8$, (a) $H = H_{c} - 0.4$, (a) $H = H_{c}$, where $H_{c}$ is the coercive field where the average magnetisation crosses from negative to positive.}
    \label{snaps}
\end{figure}

Building on this insight, it is natural to consider \emph{athermal quasistatic driving} (AQD), where one starts with large negative $H$, so the state of minimal energy has all spins in the down state, $s_i=-1$.  The field $H$ is increased quasistatically, while locally minimizing the energy.  This can be implemented by increasing $H$ in very small steps, and performing a local energy minimization after each step. In practice, it is more efficient to identify directly the value of $H$ at which the next avalanche will occur.  Then one finds the avalanches one at a time, without having to simulate the intervening small steps in which the spin configuration remains constant.
Snapshots from an AQD simulation are shown in Fig.~\ref{snaps}, the avalanches are visible as clusters of spins have flipped together.

In this work, we go beyond AQD, and consider the effect of increasing $H$ in the RFIM at finite temperature.  The field $H$ is increased in steps of size $\Delta H$, with a time $\Delta t$ between each step (measured in MCS), so the overall rate of change of $H$ is $\Gamma = \Delta H/\Delta t$.  We refer to this numerical experiment as \emph{thermal driving}.
Taking $\Delta H \to0$ or $\Delta t \to \infty$ at fixed $T>0$ yields \emph{thermal quasistatic driving}.  This work focusses on the differences between thermal driving and AQD.  To facilitate this, we give a brief review of the AQD behaviour, which has been explored in detail, both numerically \cite{perkovic_disorder-induced_1999} and theoretically \cite{sethna_hysteresis_1993,perkovic_avalanches_1995,dahmen_hysteresis_1996}.

\subsection{Scaling behaviour for athermal quasistatic driving : a brief review}
\label{sec:aqd-review}

\newcommand{\mbar}{\overline{m}}

The magnetisation of the RFIM is
\beq
m = \frac{1}{N} \sum_i s_i \; .
\eeq
For a single realisation of disorder in AQD, write $m(H)$ for the magnetisation as a function of $H$, this is a deterministic piecewise-constant function.  Averaging over the disorder, one obtains a smooth curve $\mbar(H)$, note this depends implicitly on the system size $N$ as well as the disorder strength $R$.

The coercive field $H_c$ is defined as the field where $\mbar(H)$ passes through zero.   
The behaviour of $\mbar(H)$ differs for strong and weak disorder, separated by a critical disorder strength $R_c$.  In the limit of large systems, there is a sharp distinction between systems with $R>R_c$, where $\mbar(H)$ is a smooth curve, and $R<R_c$, where $\mbar(H)$ has a jump discontinuity at $H_c$.  This discontinuity corresponds to a macroscopic (system-spanning) avalanche.  (For the system without disorder, $H_c$ is the zero-temperature spinodal, and the macroscopic avalanche involves all spins flipping simultaneously.)  This system is athermal, but the critical behaviour near $R_c$ shares many similarities with those of equilibrium systems at critical points, including universal scaling behaviour.%\footnote{The RFIM also has its own equilibrium critical point: for sufficiently small disorder, there is an $R$-dependent critical temperature $T_c$ below which the equilibrium state becomes ferromagnetic.  However, this critical point is not important for the non-equilibrium behaviour discussed in this work.}

Within AQD, consider the probability distribution $D(S,R,H)$ of avalanche sizes that occur at a given value of $H$.  This distribution depends also on the disorder strength $R$, it is normalised as $\sum_S D(S,R,H)=1$.  Define also $n(H)$ such that the average number of avalanches occurring for fields between $H$ and $H+dH$ is $n(H)dH$.  Then one may integrate the distribution of avalanches over the field $H$ to obtain the full distribution of avalanche sizes
\beq
D^{\rm int}(S,R) = \frac{ \int_{-\infty}^\infty D(S,R,H) n(H) dH }{ \int_{-\infty}^\infty n(H) dH } \; ,
\eeq
which is again normalised as $\sum_S D^{\rm int}(S,R)=1$.  In the following, it will be useful to consider the corresponding quantity without normalisation
\beq
F^{\rm int}(S,R) =  \int_{-\infty}^\infty D(S,R,H) n(H) dH \; ,
\eeq
which gives the average number of avalanches of size $S$ that occurred during the driving procedure.
We refer to $D$ as the \emph{binned} distribution, since it is obtained numerically by counting the number of avalanches that occur within a particular range (``bin'') for the field $H$.  This may be contrasted with the integrated distribution $D^{\rm int}$.  

An important aspect of the RFIM -- which is related to the observation of Barkhausen noise in disordered magnets \cite{perkovic_avalanches_1995} -- is that these avalanche distributions exhibit characteristic scaling behaviour in the critical regime $R\approx R_c$.  Define
\beq
h =  H-H_{c} \; , \qquad r = \frac{ R_{c}-R }{ R} \; .
\eeq
Within the critical regime (corresponding to sufficiently small $h,r$),  both mean-field~\cite{dahmen_hysteresis_1996,sethna_hysteresis_1993} and three-dimensional systems~\cite{perkovic_disorder-induced_1999} obey
\begin{equation}
    D(S, R, H) \sim S^{-\tau} \mathcal{D}_{\pm}\left(S^{\sigma} |r|, h /|r|^{\beta \delta}\right) \; ,
    \label{binsca}
\end{equation}
where $\mathcal{D}_+$ and $\mathcal{D}_-$ are scaling functions, which apply for $h>0$ and $h<0$ respectively, and $\beta,\delta,\sigma,\tau$ are critical exponents.  Similarly
\begin{equation}
    D^{\rm int}(S, R) = S^{-(\tau+\delta\beta\sigma)} \mathcal{D}^{({\rm int})}_{\pm}\left(S^{\sigma} |r|\right) \; .
    \label{ints}
\end{equation}

Note that for $R=R_c$ then the integrated distribution follows a power law $D_{\rm int}(S, R_c) \sim S^{-(\tau+\delta\beta\sigma)}$.  The binned distribution follows a power law only if $(R,H)=(R_c,H_c)$, then
$D(S,R_c,H_c) \sim S^{-\tau}$.   In mean-field, the exponents for these power laws are $(\tau,\tau+\beta\delta\sigma)=(3/2, 9/4)$ while in three dimensions their numerical estimates \cite{perkovic_disorder-induced_1999} are 
\beq
(\tau,\tau+\beta\delta\sigma)=(1.60,2.03) \; ,
\eeq
valid to 3 significant figures.

Note that the scaling forms (\ref{binsca},\ref{ints}) apply in the thermodynamic limit $N\to\infty$.  In finite systems, the large-$S$ tails of $D$ and $D_{\rm int}$ must be cut off (no avalanche can have $S>N$).  In practice, individual  simulations for $R\approx R_c$ typically yield a broad distribution of avalanches, together with a single system-spanning (very large) avalanche that occurs at $H_c$.  For finite systems, this behaviour persists for $R>R_c$, even though $\lim_{N\to\infty}\mbar(H)$ does not show any jump in this regime. The same phenomenon is also observed during thermal driving, which will be discussed in Sec.~\ref{sec:ava-dist}.

\subsection{Thermal driving}
\label{sec:therm}

We now discuss the thermal driving protocol.  The system evolves by MC dynamics, starting at a large negative field $H_0$. We take $H_0=-6$, which is the zero-temperature spinodal of the pure 3d Ising model, larger negative values lead to almost identical behaviour.  After running dynamics for a time $\Delta t$, we increase the field by $\Delta H$, and this procedure is repeated until $H = 6$, at which point the vast majority of spins in the system will have already flipped.  As noted above, the field increases at rate $\Gamma = \Delta H / \Delta t$.  It is convenient to define a rescaled time variable
\beq
\tilde{H}_t = (t/\Gamma) + H_0 \; ,
\label{equ:tH}
\eeq
which is a smoothly increasing quantity that is approximately equal to the field $H$ at time $t$.   We write $\langle \cdot\rangle$ for an average over the thermal noise, at fixed disorder.  For example $\langle m(t)\rangle$ is the average time-dependent magnetisation during thermal driving, for a single realisation of the random fields $h_i$.

The avalanches that occur at $T=0$ are no longer sharply defined for $T>0$.  However, many of the physical features of zero-temperature relaxation can be identified in thermal driving.  Motivated by this fact, we use the following algorithm to identify avalanches during thermal driving.  The definition has some arbitrariness, but it is consistent with the zero-temperature definition, and we will find that it provides useful comparisons between thermal and athermal dynamics.

Write $\bm{s}_{\rm b}$ (``before'') for the configuration of the system just before the field $H$ is increased, and $\bm{s}_{\rm a}$ (``after'')  for the configuration a time $\Delta t$ later (which is just before the next increase of $H$).  We find the set of spins which are up in $\bm{s}_{\rm a}$ and down in $\bm{s}_{\rm b}$.  Among these spins, we detect spatially-connected clusters.  Each such cluster is identified as a (finite-temperature) avalanche.   

Clearly, the identification of such clusters requires that the model has spatial structure, they do not exist in mean-field systems such as the RFIM on a fully-connected graph (sometimes referred to as an infinite-dimensional system). In fact, it is interesting that while mean-field and finite-dimensional RFIMs have very similar {avalanche scaling} at $T=0$~{\cite{sethna_hysteresis_1993, perkovic_disorder-induced_1999}}, their behaviour under thermal driving is completely different.  Specifically, it is easy to see that the only energy barriers in the energy landscape of the mean-field RFIM have height $O(1/N)$, so they are all irrelevant for the dynamics at temperatures $T=O(1)$.  In practice this means that the physics of avalanches if absent for thermal driving of the RFIM. {In addition, taking the mean field free energy of the RFIM \cite{schneider_random-field_1977} and finding its minima exactly reproduces the self-consistency condition of the magnetisation for the athermal driven system \cite{sethna_hysteresis_1993}, implying that the mean field RFIM is always in its thermodynamic stable or metastable state even in the athermal limit. This fact can be further confirmed by the lack of hysteresis of the magnetisation curve of the mean field athermal RFIM when driven with an external field at disorders at and below the critical value.} However, this is not the case in finite dimensions, as we shall see below.

Setting $T=0$ in this algorithm recovers the avalanches of AQD dynamics, as long as $\Delta t$ is large enough  for each avalanche to finish before the next step in $H$, and $\Delta H$ is small enough that different avalanches in the same step remain spatially separated.  On the other hand, if $T>0$ and $\Delta t$ is extremely large then the system will have enough time to fully equilibrate between each step in $H$.  This results in quite different behaviour where $\langle m(t) \rangle$ coincides with the average magnetisation of an equilibrated system in a field $H$.  For example $H_c=0$ in this latter case.  

Hence we identify two special limits: setting $T=0$ and taking very small $\Gamma\to0$ recovers AQD, while setting $T>0$ and taking $\Gamma\to0$ gives quasistatic thermal driving.  In this latter case, the notion of an avalanche is not useful.  One sees the difficulties of measuring avalanches at finite temperature: one must take $\Gamma$ in an intermediate range that is slow enough to mimic AQD, but large enough to avoid the quasistatic thermal limit.  

In practice, the appopriate range of $\Gamma$ is quite large.  This is due to activated dynamic scaling, as proposed by Fisher \cite{fisher_scaling_1986} and Villain \cite{villain_nonequilibrium_1984}.  According to this theory, near-critical relaxation proceeds by flipping {correlated} domains, and the the relaxation time for a domain of size $l$ scales as
\begin{equation}
    \ln \tau_{l} \sim \frac{l^{\theta}}{T} \; ,
    \label{equ:activated}
\end{equation}
where $\theta$ is the critical exponent {that controls the RG scaling of the random field~\cite{grinstein_ferromagnetic_1976,bray_scaling_1985}; its value is $1.49$ \cite{fytas_critical_2013} in  three dimensions}.  The thermal quasistatic limit requires $\tau_l \Gamma\ll 1$ for cluster sizes $l$ of the same order as the correlation length $\xi$.   At low temperatures, this requires an exponentially small rate $\Gamma$, so it is straightforward in practice to find rates that are sufficiently slow  to mimic AQD, but not so slow that one reaches the quasistatic thermal limit.

%Setting $T=0$ in this algorithm recovers AQD, as long as (i) $\Delta t$ is always large enough for the system to find its local energy minimum, after each step in the field; and (ii) $\Delta H$ is small enough that 

\begin{figure}[t]
    \centering
    \includegraphics[width=1\linewidth]{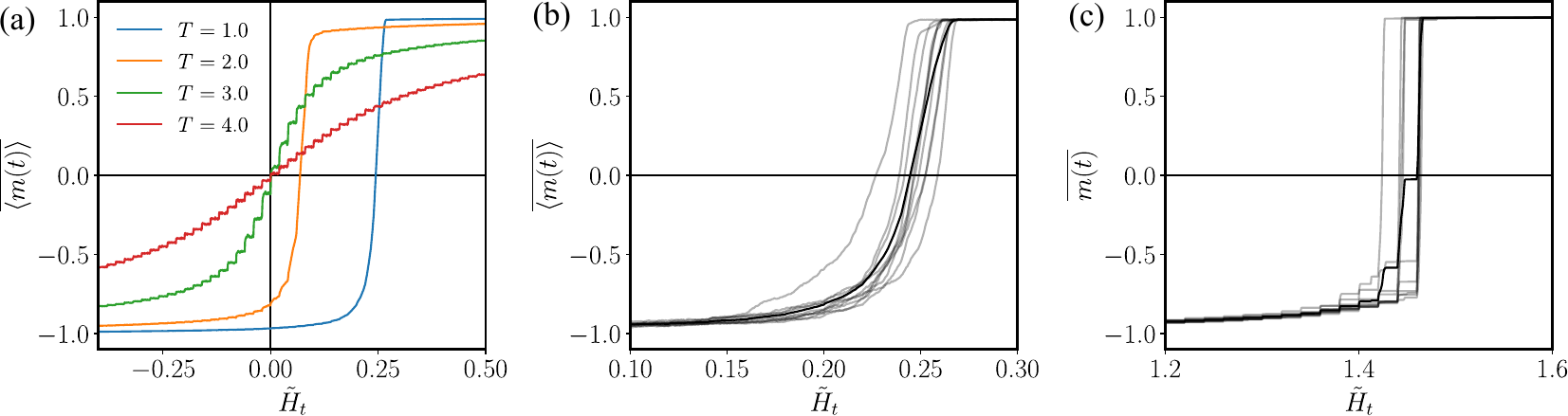}
    \caption{Time evolution of the magnetisation, plotted parametrically against the field $\tilde{H}_i$, for the 3d RFIM with $R = 2.2$, $\Delta t=2048$ and $\Delta H=0.02$. (a) The average magnetisation at various temperatures. (b) The disorder averaged curve and curves for 10 individual disorder realisations (faint lines) for $R = 2.2$, $T = 1.0$. (c) Similar to (b) but now for $T=0$ (note, the horizontal scale is different).}
    \label{fig:m-vs-h}
\end{figure}

\section{Thermal driving in the $3d$ RFIM}
\label{sec:results}

This Section shows numerical results for the $3d$ RFIM at $R=2.2$, which is very close to the critical value of the disorder in the athermal limit \cite{perkovic_disorder-induced_1999}.  Hence, the equilibrium behaviour of the RFIM is paramagnetic for all finite temperatures considered.
We investigate and discuss the extent to which the zero-temperature critical behaviour survives at moderately high temperatures.

Fig.~\ref{fig:m-vs-h} shows the behaviour of the magnetisation for thermal driving, plotted as a function of $\tilde{H}_t$, as defined in (\ref{equ:tH}).  For the lower temperatures $T=1$ and $T=2$, the magnetization shows a steep increase, as the system passes through the coercive field.  This is similar to the corresponding athermal ($T=0$) case although the coercive field is significantly larger in that case.  
Since these results are  qualitatively similar to AQD, we see that the influence of the $T\to0$ behaviour is not limited to very low temperatures $T\ll1$, but still survives at moderate temperatures. 
{(As already discussed in Sec.~\ref{sec:therm},
this picture is quite different from the situation in the mean-field RFIM, where any finite temperature is sufficient to wipe out the athermal picture, and one quickly crosses over to the quasistatic thermal behaviour.)}

By contrast, the behaviour for larger temperatures ($T=3$ and $T=4$) shows a more gradual crossover.  It is notable that the magnetisation reaches a stationary value, between each step in $H$.  Also, the coercive field $H_c\approx 0$. These observations both indicate that we are approaching the quasistatic thermal limit ($\Gamma\to0$ at finite $T$), where the system remains in thermal equilibrium throughout the crossover, and the dynamics is reversible (no hysteresis).

\subsection{Avalanche size distributions}
\label{sec:ava-dist}

\begin{figure}[t]
    \centering
    \includegraphics[width=1\linewidth]{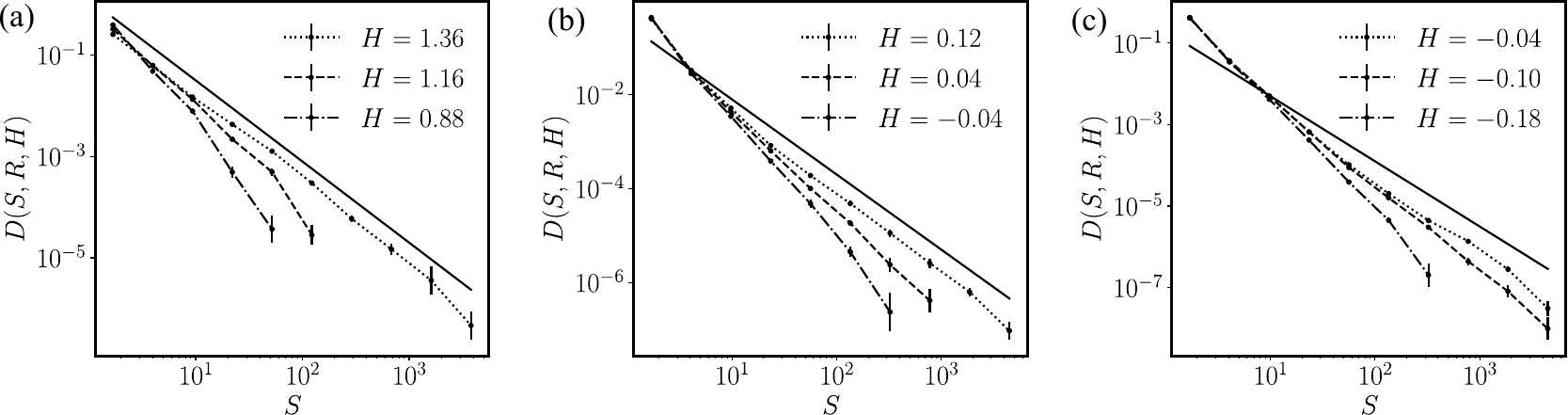}
    \caption{The avalanche size distributions of the 3d RFIM at $R = 2.2$ binned in $H$. (a) The athermal distribution. (b) Binned distribution for various external fields at $T = 1.0$, $\Delta t=2048$ and $\Delta H=0.02$ and (c) at $T = 2.0$, $\Delta t=2048$ and $\Delta H=0.02$. The solid lines correspond to the athermal critical exponent $\tau = 1.6$.}
    \label{avdist}
\end{figure}

 To further illustrate the similarity of low-temperature thermal driving and AQD, Fig.~\ref{avdist} shows the avalanche size distribution for $T=1$ and $T=2$, for fields $H$ below (but close to) the coercive field $H_c$.  For these fields, one observes power-law behaviour in the tail of $D$, that becomes increasingly well-developed as $H$ increases towards $H_c$.  Note however that
 the distribution at $H_{c}$ is dominated by a single system-spanning avalanche, as discussed earlier, so we do not show data for this case.
 %\ly{The curves are most scale-free at a value that is slightly smaller than $H_{c}$, as the distribution exactly at $H_{c}$ is dominated by the single large system-spanning avalanche or its remnants, as discussed earlier.} 
For the cases where $H$ is closest to $H_c$, the exponent of the power-law tail matches the
 %In particular, the distribution of avalanche sizes near $H_c$ shows a power-law tail with the same exponent as the 
 athermal case.  
 As discussed in Sec.~\ref{sec:aqd-review}, this power-law behaviour must be cutoff for large $S$ due to finite-size effects, and it is also difficult to obtain good statistics for very large avalanches, which are few in number.  The distributions shown here are plotted for a range of $S$ in which finite-size effects are weak.

Comparing the thermal and athermal cases in more detail, the main difference occurs for small avalanches: these are more common for finite-temperature driving.   There are two physical mechanisms that contribute to this effect.  First, the system state $\bm{s}_{\rm a}$ has thermal fluctuations, so there are clusters of spins which are transiently pointing up.  The clusters tend to be small -- they are identified as avalanches within the method, although they have no counterpart in AQD.  Second, avalanches in AQD appear when local energy minimum becomes unstable, leading to a cascade of spin flips.  The loss of stability corresponds to a vanishing energy barrier: if $T>0$ then this barrier will usually be crossed by thermal activation, before the minimum becomes unstable.  If this happens, the resulting avalanche tends to be smaller.  (One may imagine that large avalanches have less time to build up.)   This idea -- that instability thresholds are pre-empted by barrier crossing -- is related to  nucleation over small barriers, which is generically expected as spinodal instabilities are approached. It is this second effect that drives the strong decrease in $H_c$ on increasing temperature, recall Fig.~\ref{fig:m-vs-h}.
(Both effects rely on thermal activation, see also Sec.~\ref{sec:ava-scale}.)

\begin{figure}[t]
    \centering
    \includegraphics[width=1\linewidth]{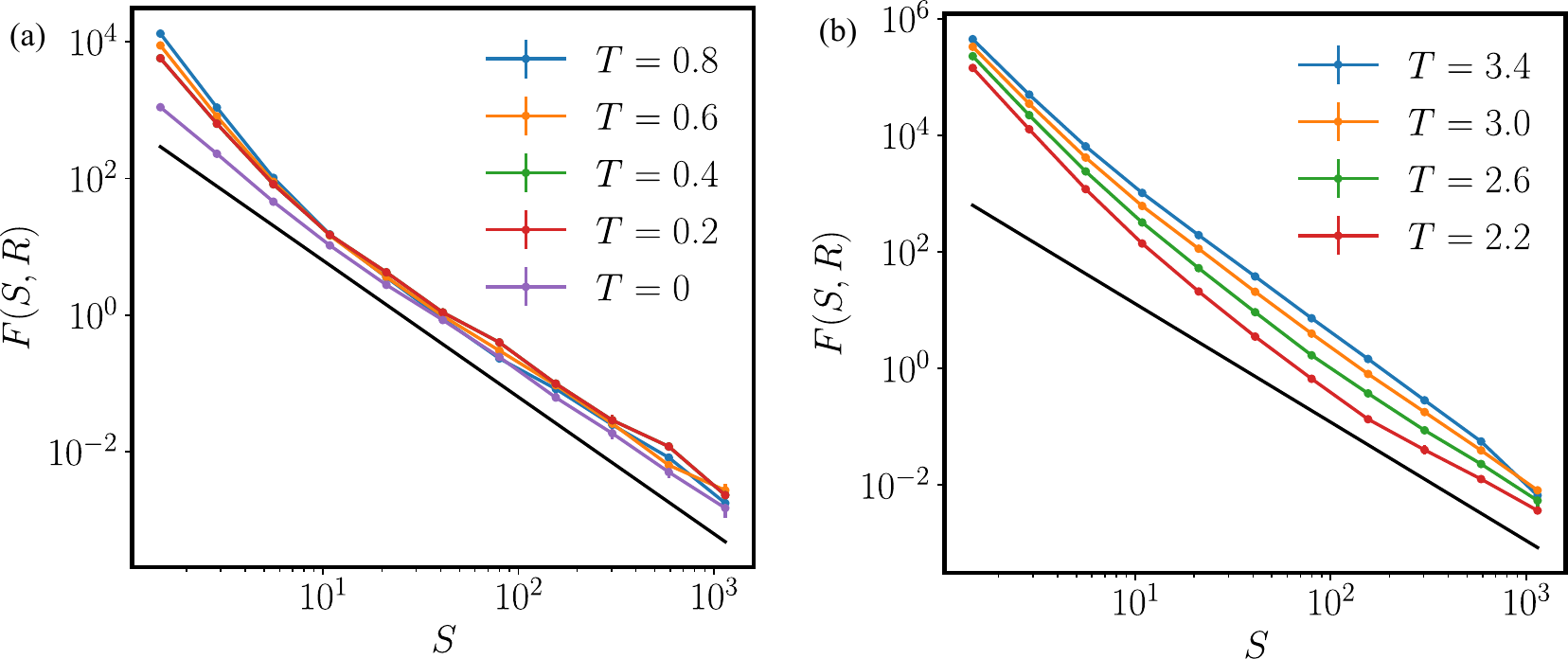}
    \caption{Integrated avalanche distribution $F(S,R)$ for the $3d$ RFIM at various temperatures, with $R = 2.2$, $\Delta t=2048$ and $\Delta H=0.02$.  
    (a) Low temperatures from $T = 0$ to $0.8$. 
    (b) Higher temperatures from $T = 2.2$ to $3.4$. The solid lines have slope $\tau+\delta\beta\sigma = 2.0$, which is the critical exponent for the athermal case.}
    \label{intdist}
\end{figure}

It is notable that the power-law behaviour of the distribution $D$ extends up to the largest avalanches plotted in Fig.~\ref{avdist}.  At $T=0$ -- and neglecting finite-size effects -- this power law should eventually be cut off, at a critical size that depends on $R-R_c$ and $H-H_c$, and diverges  at the critical point.  For $T>0$, the critical size never diverges: sufficiently close to the critical point it will depend on $T$ as well as the driving parameters $\Delta H$ and $\Gamma$.  In Fig.~\ref{avdist}, the power-law cutoffs are not seen, even at $T=2$.  Investigating them in detail would require larger system sizes, but this is not the main focus of this work.

Fig.~\ref{intdist} shows similar results for the integrated distribution of avalanches $F^{\rm int}$, normalised such that the area under the curve is their total number.  Compared to the binned distribution $D$, it is much easier to obtain good statistics for $F^{\rm int}$, as one is now sampling all avalanches that happened over the driving process.
Note that for AQD then each spin flips exactly once so $\int F^{\rm int}(S,R) S dS=N$.  For thermal driving, this number is larger, because any given spin might appear in many avalanches.

In this case we show separately the results for low temperatures $T\leq 1$ and for higher temperatures $T>2$.  For low temperatures, the distribution still shows a power-law tail for the avalanche size with the same exponent as the AQD dynamics.  This tail is visible up to $T\approx 2.5$, indicating  the zero-temperature physics is not restricted to very low temperatures.  Compared with AQD, the most striking feature is again the increased number of small avalanches, consistent with the discussion above.

For the largest temperatures ($T\geq3.0$), one also observes a reduction in the number of large avalanches, so the power-law tail in $F(S)$ is cut off. This large-$S$ cutoff signals the divergence of the thermal driving dynamics from the zero-temperature AQD behaviour.  The crossover close to $T=3$ is consistent with Fig.~\ref{fig:m-vs-h}, in which the behaviour for $T\geq3$ is consistent with \emph{thermal} quasistatic driving, which should not be expected to be similar to AQD.

We also note in passing that the behaviour of very large avalanches with $S\sim N$ also changes significantly between AQD and thermal driving.  (These large avalanches are not included in Figs.~\ref{avdist} and \ref{intdist}.)  The typical behaviour is that the single system-spanning avalanche of AQD gets fragmented into smaller avalanches, due to thermal activation.  We return to this point in Sec.~\ref{sec:fragment}, below.

\subsection{Scaling of avalanche size distribution with temperature}
\label{sec:ava-scale}

\begin{figure}[t]
    \centering
    \includegraphics[width=1\linewidth]{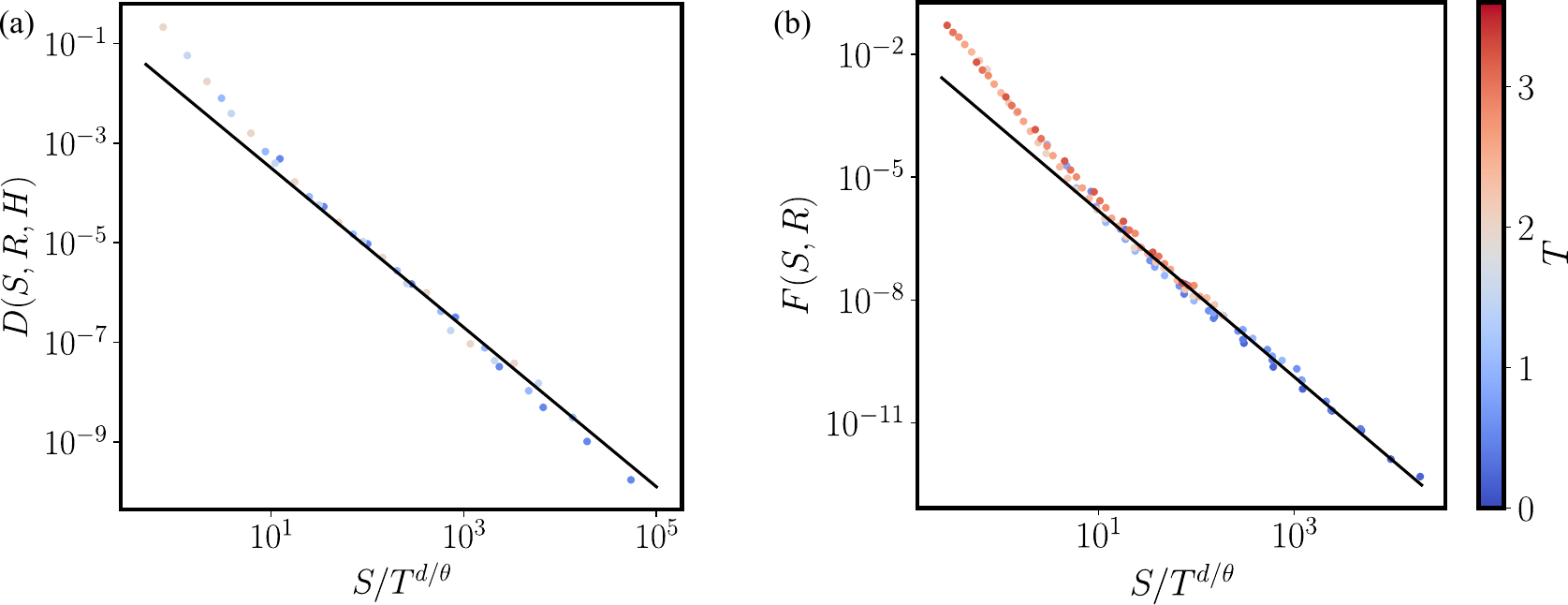}
    \caption{(a) The scaling collapse of the binned avalanche size distribution taken at the $H$ values where the distributions are the most scale invariant at temperatures  $T=(0.5,1.0,1.5,2.0)$, against the characteristic droplet size set by $T^{d/\theta}$. The solid line corresponds to the athermal critical exponent $\tau = 1.6$. (b) The scaling collapse of the integrated avalanche size distribution at various temperatures between 0.2 and 3.4, against the characteristic droplet size set by $T^{d/\theta}$. The solid line corresponds to the athermal critical exponent $\tau+\delta\beta\sigma = 2.0$.}
    \label{avn}
\end{figure}

Recall the scaling forms (\ref{binsca},\ref{ints}) for the avalanche size distribution.  For thermal driving, this distribution depends additionally on the temperature $T$ as well as the driving parameters $\Delta t$ and $\Delta H$.  The activated scaling picture of Eq.~\ref{equ:activated} \cite{fisher_scaling_1986,villain_nonequilibrium_1984} suggests the existence of a typical domain size $S_T \sim T^{d/\theta}$. Avalanches larger than $S_T$ can retain their zero-temperature behaviour, while smaller ones are easily activated by thermal fluctuations: the corresponding droplets can flip repeatedly between up and down states, causing an excess of these smaller avalanches.  
For {$(R,H) \approx (R_c,H_c)$} one may then expect $D$ and $F$ to behave as scaling functions of $S/T^{d/\theta}$.
Fig.~\ref{avn} provides evidence for this scaling.
For each temperature, we choose a value for $H$ that is close to (but smaller than) $H_c$, such that the avalanche-size distribution shows a clear power-law tail.  We then make a scaling collapse by rescaling the avalanche size by $T^{d/\theta}$.
The choice of $H$ ensures that the only $S$ scale in the plot is the thermal crossover scale at small avalanche sizes.
Hence the activated scaling theory provides a firmer theoretical explanation for the excess weight at small $S$ in the distributions of Fig. \ref{avdist} and \ref{intdist}.  
%\emph{explain how the values of $H$ were chosen.}  

Note that this scaling behaviour, which concerns the small-$S$ part of the avalanche distribution, is distinct from the behaviour for very large $S$, where one also expects scaling behaviour. This regime is described by~(\ref{binsca},\ref{ints}) for AQD, although it is also affected by thermal effects and finite-size effects, which will lead in general to power-law cutoffs as discussed in Sec.~\ref{sec:ava-dist}.  As in that section, these effects are not apparent from the moderate values of $S$ considered here.

\subsection{Dependence on dynamical parameters}

\begin{figure}[t]
    \centering
    \includegraphics[width=1\linewidth]{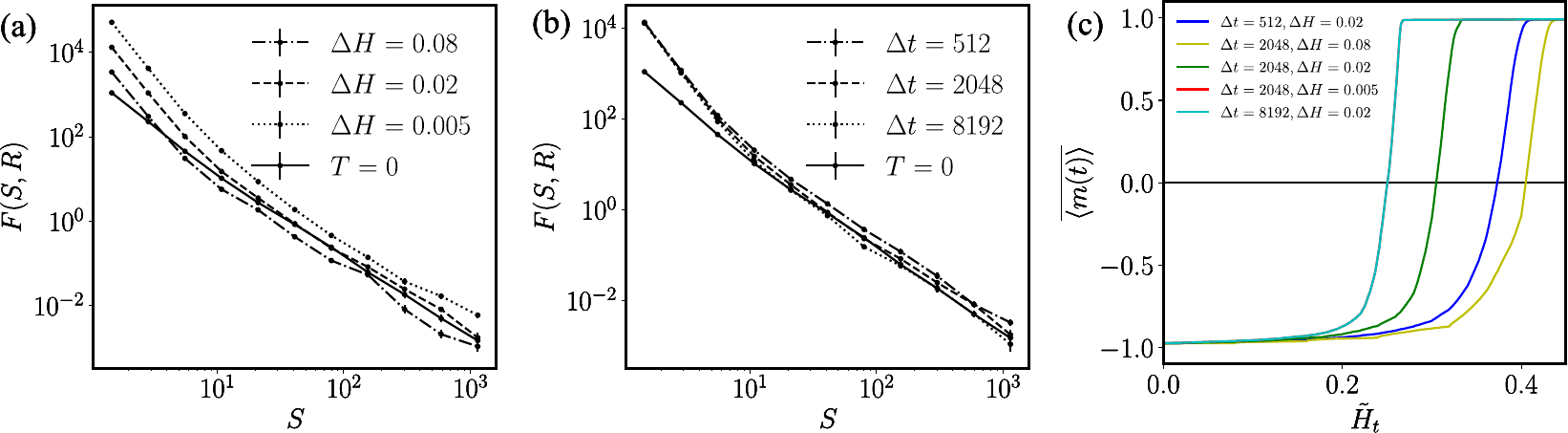}
    \caption{The integrated avalanche size distribution (normalised such that the area under each curve is equal to the total number of avalanches) at $R=2.2$ and $T=0.8$ for (a) $\Delta t = 2048$ and (b) $\Delta H = 0.02$, compared to the athermal distribution.  {(c) The disorder-averaged time series of the magnetisation at various values of $\Delta H$ and $\Delta t$. Note that the curves representing $\Delta t = 2048$, $\Delta H = 0.005$ and $\Delta t = 8192$, $\Delta H = 0.02$ completely overlap.}}
    \label{deps}
\end{figure}

As discussed in Sec.~\ref{sec:therm}, thermal driving is similar to AQD only if the driving parameters $\Delta H,\Delta t$ are chosen appropriately.  Moreover, distributions of avalanche sizes do depend quantitatively on these parameters, although the qualitative behaviour should be robust, within the appropriate range.  

This robustness is illustrated in Fig.~\ref{deps}.  In particular, the exponent of the power-law tail of $F$ remains intact of all system parameters plotted. On increasing $\Delta t$ or reducing $\Delta H$, one approaches the quasistatic thermal limit.  This leads to an increase in the number of avalanches of all sizes, because the total simulation time is larger, leaving more time for spins to flip.  On reducing $\Delta H$ at fixed $\Delta t$, one sees also a pronounced increase in the number of small avalanches.  One explanation for this is that for larger $\Delta H$, multiple small independent avalanches can overlap with each other, in which case the algorithm detects a single avalanche with a larger size.   Hence, large $\Delta H$ tends to underestimate the number of small avalanches.  
The relatively weak dependence on $\Delta t$ (at fixed $\Delta H$) is attributable to activated dynamic scaling: the relaxation time increases exponentially with the domain length scale, so moderate changes in $\Delta t$ have a weak effect on the sizes of the droplets that relax. {Thus, the driving parameters affect the range within which athermal scaling survives, but the scaling behaviour is robust within that range.}

However, we note that there exists a quasistatic limit at $\Delta H / \Delta t \to 0$, where the system follows equilibrium dynamics, and one would expect neither hysteresis nor power-law scaling in avalanche sizes. Thus, the size of the hysteresis loop (and the value of $H_{c}$) is dependent on the parameter choices. This is illustrated in Fig.~\ref{deps}(c), where the time series of the magnetisation at $T = 0.8$ are plotted for various parameter values. We clearly see that as the dynamics approach the quasistatic limit, the hysteresis loop becomes smaller.

\subsection{Thermal rounding of larger avalanches}
\label{sec:fragment}

As discussed in Sec.~\ref{sec:ava-dist}, finite-temperature avalanches tend to be smaller than their athermal counterparts, because they can nucleate before the system becomes unstable.  A natural emerging picture is that large zero-temperature avalanches tend to be fragmented into several smaller events at finite temperature.  

As a direct test for this hypothesis, we performed both AQD and thermal driving simulations for the same disorder realisations, and we compare the avalanches between them.  The overlap $Q$ between two avalanches is the number of spins that they have in common.  Given a thermal avalanche of size $S$, we find the zero-temperature avalanche with the largest overlap, and we record its size $S_0$.  We refer to this zero-temperature avalanche as the \emph{parent} of the thermal one.  For example, if the thermal avalanche is a fragment of the zero-temperature one, it is expected that $Q\approx S$. 

\begin{figure}[t]
    \centering
    \includegraphics[width=1\linewidth]{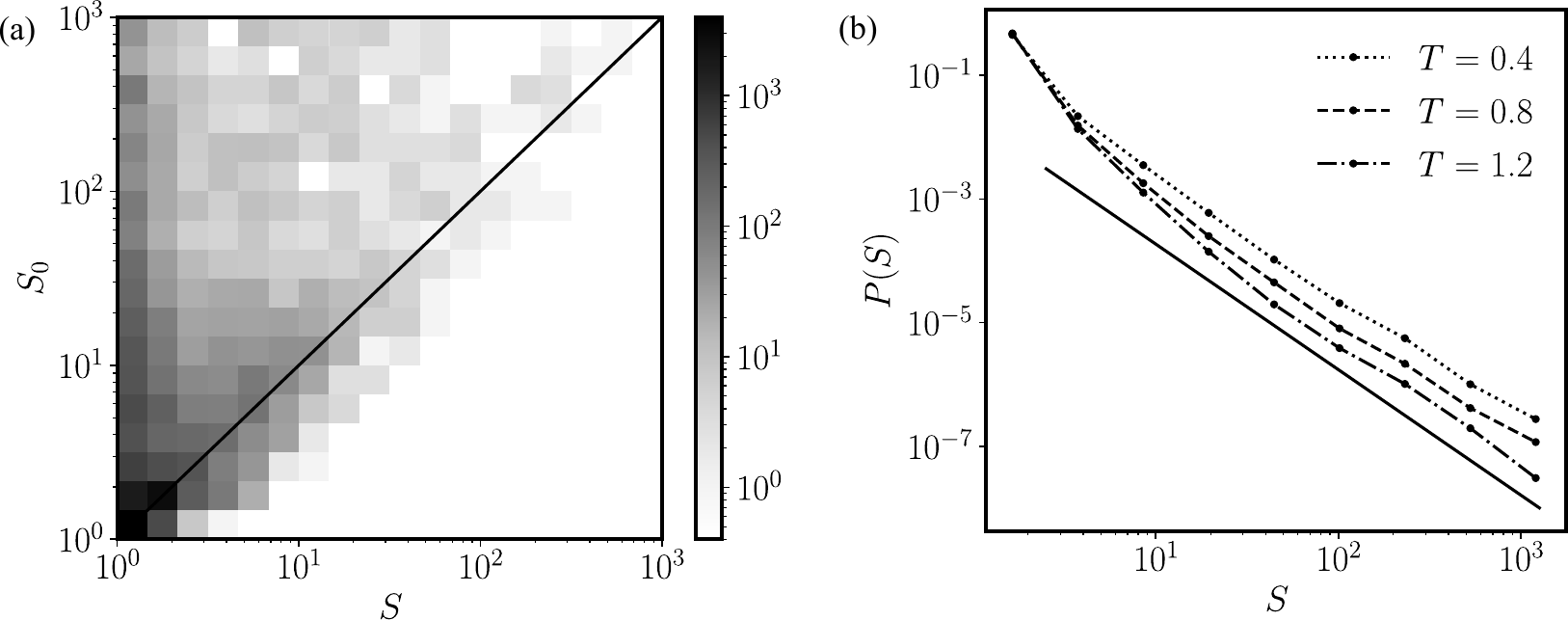}
    \caption{(a) 2d histogram of the number of thermal avalanches of size $S$ parent size $S_{0}$, at $R = 2.2$ and $T = 0.4$. The shading of the bins is in log scale. (b) The integrated size distribution of the thermal avalanches whose athermal avalanche that they share the largest number of spins with is the system spanning one. The solid line has a gradient corresponding to the exponent of the integrated avalanche size distribution $\tau+\delta\beta\sigma = 2.0$. }
    \label{fig:parent}
\end{figure}

Figure~\ref{fig:parent}(a) shows the histogram of the number of thermal avalanches of size $S$ with parent size $S_{0}$.  There is significant statistical weight for $S\approx S_0$, indicating that many  thermal avalanches can be traced to an athermal parent of similar size.
This again illustrates the robustness of the AQD picture in the presence of thermal fluctuations. %\rlj{\it do we really want to cite all of \cite{balog_criticality_2018} \cite{imry_random-field_1975} \cite{bray_scaling_1985} here?} \ly{\it ly: I would cite them if we mention the connection to the RG irrelevance of temperature, otherwise there's no need}

Also, recall from Sec.~\ref{sec:aqd-review} that a given realisation of AQD typically contains a single system-spanning avalanche that  place near $H_c$.  The parent of any thermal avalanche may be the system-spanning athermal one.  Fig.~\ref{fig:parent}(b) shows the avalanche size distribution for those thermal avalanches whose parent is system-spanning.  Remarkably, this distribution shows the same power-law tail as we found when considering  all avalanches.  In other words, the fragments of the system-spanning zero-temperature avalanche are themselves power-law distributed.  This result is interesting because it shows that the fragmentation of avalanches due to thermal fluctuations does not generate a simple cutoff in the size distribution, at large avalanches.  Instead, it creates additional avalanches that still contribute to a power law distribution, presumably due to an underlying fractal structure.  (Note that thermal driving still results in a system-spanning avalanche, so one should think of the athermal avalanche fragmenting into one system-spanning component, and a power-law distribution of smaller ones.)

\section{Conclusion and outlook}
\label{sec:conc}

{We have shown that thermal driving in the $3d$ RFIM retains many features of athermal driving.  In particular, while avalanches are no longer sharply defined at finite temperature, their signatures are still imprinted on the dynamics, in that spins flip in large clusters, for temperatures up to twice the exchange energy of the model.  The resulting clusters (or finite-temperature avalanches) can be detected by an appropriate numerical algorithm, and show scaling behaviour near the coercive field, when the disorder $R$ is close to $R_c$. {The resulting finite-temperature avalanches show a significant statistical correspondence with their athermal parents.} Also, the avalanche size distribution has power-law behaviour, with the same exponents as the athermal case.  The behaviour is also robust to the driving parameters, $\Delta t$ and $\Delta H$.
We also emphasised that this behaviour in finite-dimensional systems is distinct from that of mean-field models, which do not have avalanches at any finite temperature.}

Comparing the athermal case with driving 
at moderate temperatures, the clearest difference in behaviour is an excess of small thermally-activated avalanches, which appear below a typical size $S_{\rm th} \sim T^{d/\theta}$, as predicted by activated dynamical scaling~\cite{fisher_scaling_1986,villain_nonequilibrium_1984}.  At higher temperatures, one also observes a change in the behaviour of large avalanches, which are disrupted by thermal effects, above some typical size $S_{\rm frag}$.  The athermal phenomenology survives in an intermediate range $S_{\rm th}<S<S_{\rm frag}$, which shrinks gradually on increasing $T$.  Eventually, the two cutoffs merge, and the avalanche picture is lost.

Our results are consistent with observations of athermal avalanche behaviour in experimental systems (and computer simulations) at temperatures where thermal fluctuations cannot be neglected.  For example, we expect that experiments on crackling noise in impure magnets should produce statistics that resemble the athermal scaling laws, subject to cutoffs at high and low amplitudes (which would anyway be expected in practical setups). {Experiments have successfully established the survival of athermal avalanche size scaling in a model magnet at liquid helium temperatures and its breakdown near room temperature~\cite{xu_barkhausen_2015}, but a more detailed analysis of the crossover region between these two regimes would be welcome.
Theories which connect RFIM physics with glassy systems \cite{ozawa_random_2018, ozawa_role_2020} also rely on the survival of (qualitative) aspects of the athermal case, so these results support the relevance of such theories, even in the face of significant thermal fluctuations. 

More generally for disordered systems, our results suggest that the common practice of neglecting thermal fluctuations when modelling finite-temperature disordered systems (such as modelling the shearing of glasses as an athermal quasistatic process \cite{maloney_amorphous_2006}) can produce meaningful physical results within a wide range of temperatures, up to corrections at the extremities of the avalanche size distribution. 
The general expectation is that the crossover from athermal to equilibrium-like behaviour is controlled by a competition of a driving time scale (here $\Gamma$) and an equilbrium relaxation time.  This is consistent with our results, but we emphasize that this crossover affects different parts of the avalanche size distributions in different ways, which is partly attributable to the broad distribution of length and time scales underlying equilibrium relaxation in this model.

We note that there have being recent theoretical advancements in the understanding of the finite temperature crossover between the depinning and equilibrium universality classes of driven elastic interfaces in disordered environments \cite{https://doi.org/10.48550/arxiv.2201.12652}, a system similar to ours. That work analyses a particular two-point correlation function which exhibits a cusp, whose properties are related to the existence of avalanches. This enables a connection with the theory of the functional renormalisation group. Their work implies a temperature-dependent cutoff on the tail of the avalanche size distribution, consistent with our results.

Looking forward, it would be interesting to investigate other aspects of RFIM dynamics at finite temperature.  For example, theories of coupled replicas relate static properties of glasses to RFIM criticality~\cite{franz_universality_2013}, as supported by numerical results~\cite{guiselin_random-field_2020,jack_phase_2016}.  It has been proposed that responses of such materials to temperature changes is related to nucleation and growth dynamics~\cite{jack_melting_2016,bouchaud_adam-gibbs-kirkpatrick-thirumalai-wolynes_2004} for the RFIM, but this process has been rather little studied so far (see however \cite{mandal_nucleation_2021, PhysRevE.95.052140}).  From the perspective of critical phenomena, the relevance of the quenched disorder in such systems indicates that it may have strong (and potentially universal) implications for dynamical behaviour, and the results presented here indicate that this can survive up to moderate temperatures.  We look forward to future work in this direction.

\section*{References}
\typeout{}
\bibliographystyle{unsrt}
\bibliography{bibliography}
\end{document}